# Diagnosis of Switching Systems using Hybrid Bond Graph

Taher MEKKI, Slim TRIKI and Anas KAMOUN

Research Unit on Renewable Energies and Electric Vehicles (RELEV), University of Sfax
Sfax Engineering School (ENIS), Tunisia
*Taher.mekki@gmail.com*
*Slim.triki1@gmail.com*
*Anas.kamoun@enis.rnu.tn*

**Abstract**
Hybrid Bond Graph (HBG) is a Bond Graph-based modelling approach which provides an effective tool not only for dynamic modeling but also for fault detection and isolation (FDI) of switching systems. Bond graph (BG) has been proven useful for FDI for continuous systems. In addition, BG provides the causal relations between system's variables which allow FDI algorithms to be developed systematically from the graph. There are many methods that exploit structural relations and functional redundancy in the system model to find efficient solutions for the residual generation and residual evaluation steps in FDI of switching systems. This paper describes two different techniques, quantitative and qualitative, based on common modelling approach that employs HBG. In quantitative approach, global analytical redundancy relationships (GARRs) are derived from the HBG model with a specified causality assignment procedure. GARRs describe the system behaviour at all of its operating modes. In qualitative approach, functional redundancy can be captured by a Temporal Causal Graph (TCG), a directed graph that may include temporal information.

**Keywords:** *Hybrid Bond Graph (HBG), Global Analytical Redundancy Relation (GARR), Temporal Causal Graph (TCG).*

## 1. Introduction

Several physical systems with switching are nonlinear dynamic systems. When switching occurs, the system may change its mode of operation. If a system has $n$ switching states, then it gives rise to $2^n$ possible operating modes. One way to represent mode switching is to generate $2^n$ sets of differential-algebraic equations (DAEs). Each set describes continuous behaviour of system in that particular mode. In practice, not all modes are practically realizable. Many recent researches on switching systems have been devoted to the synthesis of control laws that guarantee not only the stability but also good performances [1]. The control algorithms are generally developed considering that the system works in normal situation. Unfortunately, when failures occur, these algorithms become inefficient and even dangerous for the system itself or its environment. In order to reach higher performances and more rigorous security specifications, a Failure Detection and Isolation system has to be implemented. The literature in that field is abundant and different solutions have been proposed for continuous or discrete, linear and non linear systems. However, only few solutions have been proposed for switching systems.

Traditionally, two different communities: (1) the Systems Dynamics and Control Engineering (FDI) community (e.g., [2] and [3]), and (2) the Artificial Intelligence Diagnosis (DX) community (e.g., [4] and [5]), have developed model-based diagnosis approaches. The two communities have employed different kinds of models, and made different assumptions concerning robustness of the generated solution with regard to modeling errors, measurement noise, and disturbances. The general principle of all model-based FDI approaches is to compare the expected behavior of the system, given by model, with its actual behavior. The first step of a FDI procedure consists in generating a set of residuals. These residuals are special signals that reflect the discrepancy between the two behaviors. Analytic Redundancy Relation (ARR) methods are classically used for residual generation in the FDI community [6]. The DX community has developed methods such as possible conflicts [7, 8], and analysis of temporal causal graphs [9, 10] for diagnosis of continuous systems. These methods are based on the structural analysis of dynamic models, much like the ARR schemes developed by the FDI community. The two communities use different algorithms, but the overall framework for fault isolation is similar, defined by a two-step process: (i) residual generation, followed by (ii) residual evaluation [2, 3].

In this work, we focus on the Hybrid Bond Graph as unified graphical method of modeling and diagnosis of switching systems. There are two main approaches while using HBGs: those who use switching elements with fixed causality [11, 12], and those who use ideal switching elements which changing causality [13]. Therefore, we start with common modeling framework, hybrid bond graph, to describe and compare the ARR approach developed by [14] with temporal causal graph based diagnosis [15, 16]. In particular, the residual generation and evaluation algorithms used by the two methods are presented and a discussion between the algorithms is established.

# 2. Quantitative hybrid bond graph-based fault detection and isolation

## 2.1 Analytical Redundancy Relations (ARRs) and Global ARRs

An ARR obtained from a physical law represents some conservation phenomena: Bernoulli equation etc. in hydraulic domain; Newton's law etc. in mechanical domain; and Kirchoff's law etc. in electrical domain. The ARRs for ordinary (non hybrid) system can be derived algorithmically from derivative causality bond graph model or Diagnostic Bond Graph (DBG) [17], [18]. Whereas the bond graph model for hybrid system with discrete mode changes is called the Hybrid Bond Graph, and the GARRs are derived in similar manner from differentially causalled HBG or Diagnostic Hybrid Bond Graph (DHBG). A DHBG is a HBG which describes the behavior of a hybrid system at all modes based on unified set of causality assignment [19]. In other words, for a given DHBG, a reassignment of causality to the HBG is not required to describe the behavior of the system at different operating modes. This unique feature of DHBG implies that the causal paths of graph maintain the same structure, except that some of the sub-paths are eliminated due to the OFF states of the controlled junctions. Based on these uniform causal paths, a set of unified relations is derived to describe the hybrid system at all modes. This set of relations which is called Global ARRs (GARRs), are utilized as ARRs for the hybrid systems. Likewise for continuous system, the GARRs provide a convenient tool to deduce the fault detectability and isolability of a hybrid system.

To illustrate how to derive the GARR for FDI application, consider the three coupled tanks depicted in Fig. 1. These tanks are connected by pipes which can be controlled by different valves. Water can be filled into the left and right tanks using two identical pumps. Measurements available from the process are the continuous water levels $h_i(t)$ of each tank. The connection pipe, with valve $R_{12}$ (res. $R_{23}$), between the tank 1 and 2 (res. 2 and 3) is placed at a height of 0.5m (res. 0.7m).

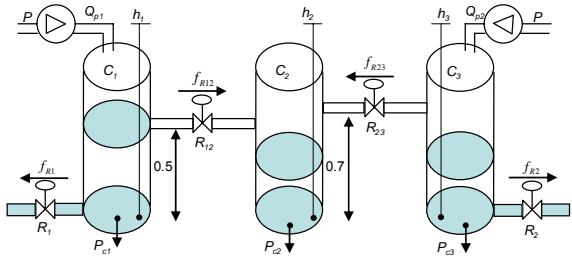

Fig. 1. Three Tank System.

Fig. 2 shows the hybrid bond graph model of this system. The two flow sources into tanks 1 and 3 are indicated by $Sf_1$ and $Sf_2$, respectively. The tank capacities are shown as $C_1$, $C_2$ and $C_3$. The pipes are modeled by resistances $R_1$, $R_{12}$, $R_{23}$ and $R_2$. Pumps and valves are modeled by controlled junctions, which are shown in the figure as junctions with subscripts ($1_1$, $1_2$, $1_5$, and $1_6$). The control signals for turning these junctions *on* and *off* are generated by the finite state automata in Fig. 2. For autonomous transitions in the system, also modeled by controlled junctions, the transition conditions computed from system variables (junctions $1_3$ and $1_4$). A mode in the system is defined by the state of the six controlled junctions in the hybrid bond graph model. Therefore, theoretically the system can be in $2^4$ different modes.

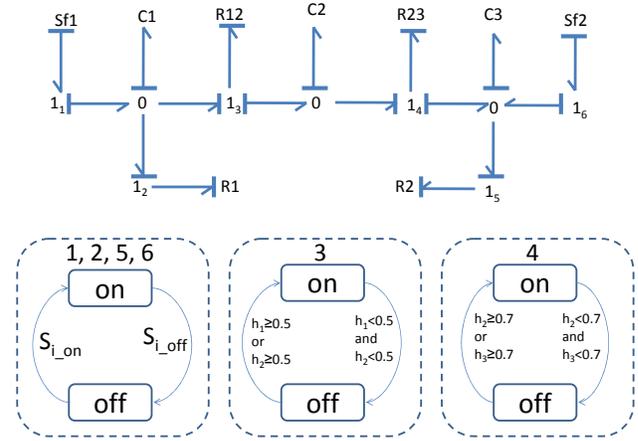

Fig. 2. HBG of the three tank system.

Fig. 3 shows the DHBG model of the hybrid tank system deduced from the three step procedure Hybrid Procedure Assignment Causality Sequential (SCAPH) and Model Approximation (MA) detailed in [19]. First step, a $R_s$ component is added to every sensor junction; in this example a very-high resistive component is added to the junction $0_1$, $0_2$ and $0_3$. The second step is to apply the SCAPH algorithm. In this step, all controlled junction are assigned with their preferred causality. As shown in Fig. 3, the output variables of the controlled-junctions $1_{c1}$ and $1_{c2}$ ($e_5$ and $e_{10}$, respectively) are assigned as inputs of the 1-port component ($R_{12}$ and $R_{23}$, respectively). There is no source connected to the controlled-junction. Then the two sources $Q_{p1}$ and $Q_{p2}$ are assigned with their preferred causality. Since the DHBG is required to generate the GARR, the three storage components $C_1$, $C_2$ and $C_3$ are assigned with their preferred derivative causality. We extend the causal implication using the components constraints to remain bonds to complete the SCAPH algorithm. The final step is to eliminate every $R_s$ that is in indifferent causality. In our case $R_s$ added to the junction $0_1$ and $0_3$ are redundant and, therefore, are removed from the DHBG model of Fig. 3.

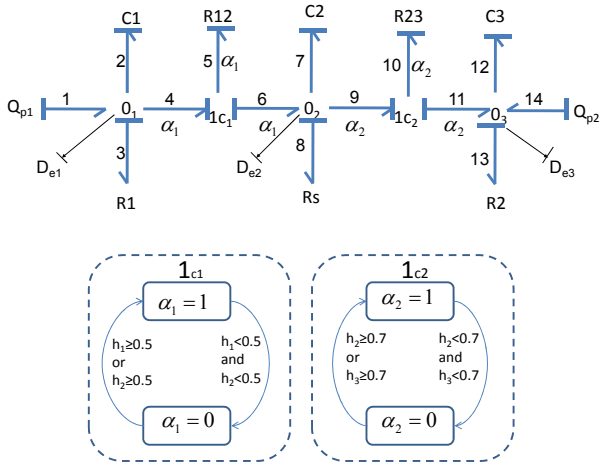

Fig. 3. DHBG of the three tank system.

The constitutive relation of the $R_s$ component connected to the junction $0_2$ is $f_8 = 0$. From the DHBG, the constitutive relations of junctions $0_1$, $1_{c1}$, $0_2$, $1_{c2}$ and $0_3$ are given by the following equations:

Junction $0_1$
$$f_1 - f_2 - f_3 - \alpha_1 f_4 = 0 \quad (1)$$
Junction $1_{c1}$
$$e_4 - e_5 - e_6 = 0 \quad (2)$$
Junction $0_2$
$$\alpha_1 f_6 - f_7 - f_8 - \alpha_2 f_9 = 0 \quad (3)$$
Junction $1_{c2}$
$$e_9 - e_{10} - e_{11} = 0 \quad (4)$$
Junction $0_3$
$$\alpha_2 f_{11} + f_{14} - f_{12} - f_{13} = 0 \quad (5)$$
Where
$$f_4 = f_6 = \begin{cases} 0 & \text{if } h_1 < 0.5 \text{ and } h_2 < 0.5 \\ \frac{1}{R_{12}}(De_1 - 0.5) & \text{if } h_1 > 0.5 \text{ and } h_2 < 0.5 \\ -\frac{1}{R_{12}}(De_2 - 0.5) & \text{if } h_1 < 0.5 \text{ and } h_2 > 0.5 \\ \frac{1}{R_{12}}sign(De_1 - De_2)*|De_1 - De_2| & \text{if } h_1 > 0.5 \text{ and } h_2 > 0.5 \end{cases}$$

$$\Rightarrow f_4 = f_6 = \frac{\alpha_1}{R_{12}} sign(\max(De_1,0.5) - \max(De_2,0.5))*(\max(De_1,0.5) - \max(De_2,0.5))$$

$$f_9 = f_{11} = \begin{cases} 0 & \text{if } h_2 < 0.7 \text{ and } h_3 < 0.7 \\ \frac{1}{R_{23}}(De_2 - 0.7) & \text{if } h_2 > 0.7 \text{ and } h_3 < 0.7 \\ -\frac{1}{R_{23}}(De_3 - 0.7) & \text{if } h_2 < 0.7 \text{ and } h_3 > 0.7 \\ \frac{1}{R_{23}}sign(De_2 - De_3)*|De_2 - De_3| & \text{if } h_2 > 0.7 \text{ and } h_3 > 0.7 \end{cases}$$

$$\Rightarrow f_9 = f_{11} = \frac{\alpha_1}{R_{12}} sign(\max(De_2,0.7) - \max(De_3,0.7))*(\max(De_2,0.7) - \max(De_3,0.7))$$

Three structurally independent GARRs can be generated from (1), (3) and (5) after eliminating the unknown variables. The GARRs are obtained as follows:

$$GARR1 = Q_{p1} - C_1\frac{d}{dt}De_1 - \frac{1}{R_1}De_1$$
$$- \frac{\alpha_1}{R_{12}}sign(\max(De_1,0.5) - \max(De_2,0.5))*(\max(De_1,0.5) - \max(De_2,0.5)) \quad (6)$$

$$GARR2 = \frac{\alpha_1}{R_{12}}sign(\max(De_1,0.5) - \max(De_2,0.5))*(\max(De_1,0.5) - \max(De_2,0.5))$$
$$-C_2\frac{d}{dt}De_2 - \frac{\alpha_2}{R_{23}}sign(\max(De_2,0.7) - \max(De_3,0.7))*(\max(De_2,0.7) - \max(De_3,0.7)) \quad (7)$$

$$GARR3 = \frac{\alpha_2}{R_{23}}sign(\max(De_2,0.7) - \max(De_3,0.7))*(\max(De_2,0.7) - \max(De_3,0.7))$$
$$+Q_{p2} - C_3\frac{d}{dt}De_3 - \frac{1}{R_2}De_3 \quad (8)$$

2.2 Fault detectability and fault isolability

Unlike continuous systems, hybrid systems are multiple modes in nature. This feature suggests that the system's fault detectability and fault isolability are required to be evaluated at different operating modes for effective FDI analysis and designs. The unified characteristic of the GARRs provide a convenient way to generate the Fault Signature Matrix (FSM) of each operating mode. Table 1 shows the FSM of the three tank system and table 2 describes the modes. The fault detectability and fault isolability of each parameter is gained from the $\{D_b, I_b\}$ values of the four FSMs. This information can be summarized in table 3.

Table 1: FSM for the three tank system

|  | GARR1 | GARR2 | GARR3 | $D_b$ | $I_b$ |
|---|---|---|---|---|---|
| $R_1$ | 1 | 0 | 0 | 1 | 0 |
| $C_1$ | 1 | 0 | 0 | 1 | 0 |
| $R_{12}$ | $\alpha_1$ | $\alpha_1$ | 0 | $\alpha_1$ | $\alpha_1$ |
| $C_2$ | 0 | 1 | 0 | 1 | 1 |
| $R_{23}$ | 0 | $\alpha_2$ | $\alpha_2$ | $\alpha_2$ | $\alpha_2$ |
| $C_3$ | 0 | 0 | 1 | 1 | 0 |
| $R_2$ | 0 | 0 | 1 | 1 | 0 |

Table 2: Modes of the system

| Modes  | $\alpha_1$ | $\alpha_2$ |
|--------|------------|------------|
| Mode 1 | 1          | 1          |
| Mode 2 | 1          | 0          |
| Mode 3 | 0          | 1          |
| Mode 4 | 0          | 0          |

Table 3: Fault Detectability and fault Isolability of components

| $\theta$ | Detectability | Isolability |
|----------|---------------|-------------|
| $R_1$    | all-mode      | Nil         |
| $C_1$    | all-mode      | Nil         |
| $R_{12}$ | Mode 1, 2     | mode 1, 2   |
| $C_2$    | all-mode      | all-mode    |
| $R_{23}$ | Mode 1, 3     | mode 1, 3   |
| $C_3$    | all-mode      | Nil         |
| $R_2$    | all-mode      | Nil         |

In this work, MATLAB 7.0 is used to simulate the model of the tank system. The parameters of the system are fixed as follow; $R_1 = R_{12} = R_{23} = R_2 = 1 m^{-1} s^{-1}$, $C_1 = C_2 = C_3 = 1 kg^{-1} m^4 s^2$, and simulation time is fixed to 10$s$ with sampling time $t_s$=0.01s. The inputs $Q_{p1}(t)$, $Q_{p2}(t)$ are presented in Fig. 4.

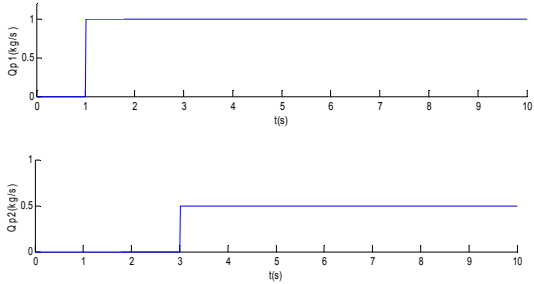

Fig. 4. Inputs on the system.

A fault is simulated in $R_{12}$ component where its parametric value changes abruptly from 1 to 5 at $t$=1$s$. The fault profile is shown in Fig. 5. Figure 6 depicts the measured variables and the switches signals.

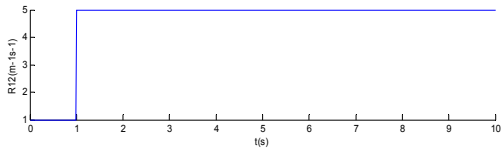

Fig. 5. Fault profile in $R_{12}$ component.

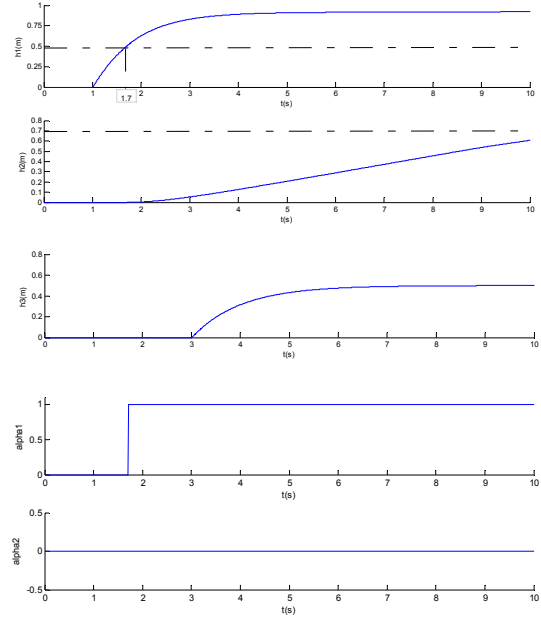

Fig. 6. plot of measured variables and operating modes alpha 1 and 2.

The residuals GARR1, GARR2 and GARR3 due to the fault in $R_{12}$ are shown in Fig. 7 where line denotes the thresholds are $\varepsilon_1 = \pm 0.02, \varepsilon_2 = \pm 0.01$ and $\varepsilon_3 = \pm 0.01$.

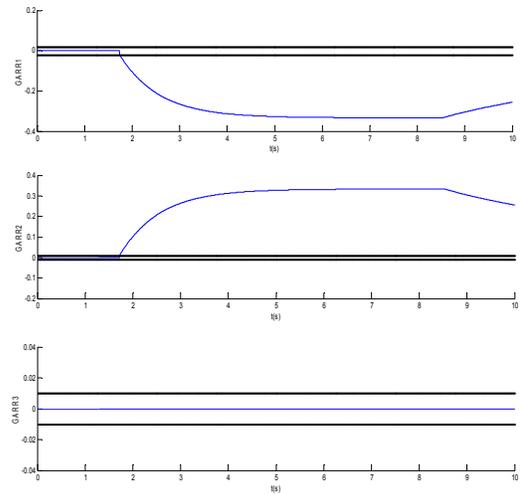

Fig. 7. Residuals responses for single fault in $R_{12}$ component

In general, if the residual exceeds the predetermined threshold, system is considered as faulty. According to the FSM tables, we can easy deduce that the faults in $R_{12}$ initiate at Mode 4, i.e. $\alpha_1 = 0, \alpha_2 = 0$, which is a non-observable mode. Hence the fault can not be detected until the system enters a mode in which fault is detectable, i.e. at time $t$=1.7s (see Fig. 6) when the system change to the Mode 2 ($\alpha_1 = 1, \alpha_2 = 0$). Fig. 7 reveals that GARR1 and

GARR2 are sensitive to fault. From the FSM table at that mode, $R_{12}$ has fault signature [1 1 0]. According to the FSM table 3, $R_{12}$ is not detectable at Mode 3.

## 3. Qualitative hybrid bond graph-based detection and isolation

### 3.1 Temporal Causal Graph and Parametrized Causality

The DX community from the field of Artificial Intelligence, have developed a number of diagnosis algorithms based on consistency-based techniques [5]. There are many works in the literature that use the BG and HBG modeling language for this purpose. For example, [9] developed a diagnosis schema, which called TRANSCEND, based on the qualitative analysis of fault transient information for diagnosis of continuous systems. Since hybrid systems cannot be described by a single continuous or discrete event model, [20] extend the TRANSCEND system to Hybrid TRNASCEND where the diagnosis algorithms use a combination of qualitative and quantitative reasoning mechanisms. [21] propose a qualitative event-based approach to fault diagnosis of hybrid systems that extends the TRANSCEND and Hybrid TRANSCEND methodologies to incorporating discrete faults.

In all proposed frameworks, system uses a *Temporal Causal Graph* (TCG) model in order to analysis fault transients. Then, the diagnosis is based on this analysis, where observed deviations from nominal behavior expressed in qualitative form are compared against qualitative predictions of faulty behavior, i.e., *fault signature*, to isolate faults. For a particular mode, the TCG is constructed based on the system equations for that mode. Using the bond graph model this process is made easy. First, causality is assigned using SCAP [22], or, in the case of HBGs, may be reassigned based on the assignment of the previous mode using Hybrid SCAP [22, 23, 24]. After that, the bonds are converted to system variables and the bond graph elements are converted into labeled edges connecting the variables of their associated bonds (see Fig. 8). Signal links are converted into single edges, with the qualitative label corresponding to the qualitative relationship between the variables.

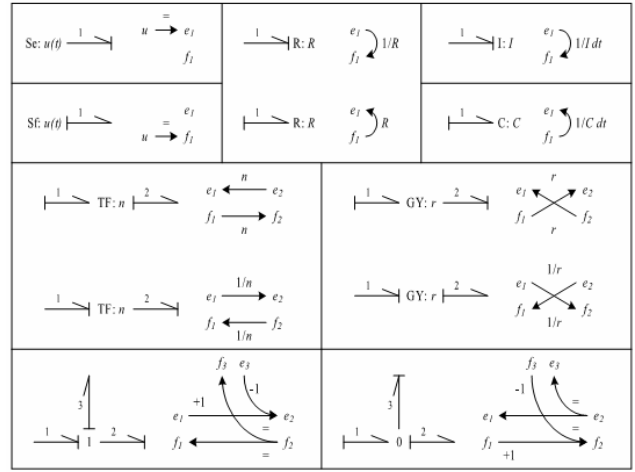

Fig. 8. Temporal Causal Graph transformations.

Fig. 9 shows a bond graph of the three-tank system and its corresponding TCG in the case when control junctions are both in mode OFF. The numbered bonds are converted to corresponding variables with subscript indicating the bond number. For example, bond 3 becomes $e_3$ and $f_3$. The resistor relates these two variables, i.e., $f_3 = \frac{1}{R_1} e_3$. The causality of the bond indicates that the 0-junction is imposing effort on $R_1$, and $R_1$ is imposing flow on the 0-junction. Therefore, the causal relationship is from $e_3$ to $f_3$. The label is $R_1$, which corresponds to the constituent equation of the resistor in the given causality. For the capacitance, the relationship between $f_2$ and $e_2$ is an integration, hence the *dt* label. Junctions sum one type of variable according to the bond signs, and set the other type equal. For the 0-junction, bond 2 is determining the effort of the junction, so $e_1$ and $e_3$ are set equal to $e_2$. According to the bond signs, $f_2 = f_1 - f_3$. Since $f_2$ must be determined, $f_1$ and $f_3$ causally affect $f_2$ with labels +1 and -1, respectively.

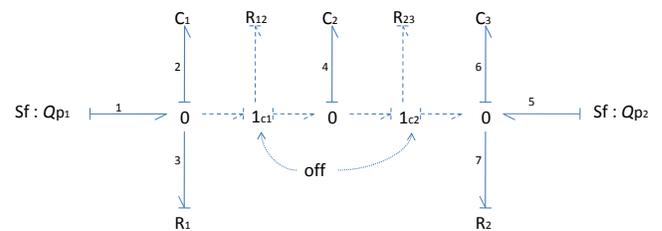

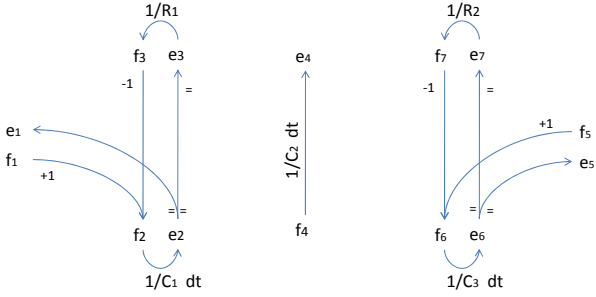

Fig. 9. Temporal Causal Graph of Three Tank System in mode 00.

In order to deal with the change of causality, the TCG can be derived for each possible system configuration or mode. However, in case of many locally acting switches, the combinatorial explosion quickly leads to an intractable problem. These problems can be mitigated to some extent by dynamically generating the TCG of each possible system mode in response to a failure. This may still result in a problem with large computational complexity which can be further reduced by measuring system variables that indicate specifically which local switches may have occurred [25] and predictions for each of the variables that determine different causal assignments are required to be made and analyzed. Once a set of possible TCGs is available, Gaussian decision techniques have been applied to compute the most likely mode of continuous behavior [9]. Others attention to hybrid diagnosis [26] concentrates on efficiently processing a set of TCGs. [27] describes how a hybrid model can be made amenable to the diagnosis algorithms that were developed in [28] by systematically generating one parametrized TCG. In this graph, the directed links are enabled by conditionals that correspond to the mode in which these links are present. The result is a set of predictions that are parametrized by the state of the local switches and the diagnosis problem then becomes one of constraint satisfaction [16]. The solution to this constraint satisfaction problem contains the possible parameter changes (i.e., the faults) and the effect on the system mode that this is required to have.

### 3.2 Parametrized Causality: Temporal Causal Matrix

To derive qualitative predictions, the system may be written as a directed graph that captures the causal (directed) relations between system variables [27]. Hence, the TCG can be represented by a weighted adjacency matrix where the columns are cause and rows are the effect variables and the entries capture the parameters on the graph edges. This is called the Temporal Causal Matrix (TCM). In the case of switched systems, modeled discontinuities result in causal changes. Therefore, the TCM may take several different forms and so do the corresponding predictions of future behavior, depending on whether a mode change occurs. To handle the change in TCM, the causal relations can be parametrized to make them dependent on the mode of operation. To this end, first the system is described in a noncausal form by using implicit equations. An implicit model of the three-tank (see Fig. 2) consists of the following equations:

$$\begin{cases} \alpha_1(-P_{c1} + P_{R12} + P_{c2}) + (1-\alpha_1)f_{R12} = 0 \\ \alpha_2(-P_{c2} + P_{R23} + P_{c3}) + (1-\alpha_2)f_{R23} = 0 \\ f_{c1} - Q_{p1} + f_{R1} + f_{R12} = 0 \\ f_{c2} - f_{R12} + f_{R23} = 0 \\ -f_{c3} + f_{R23} - f_{R2} + Q_{p2} = 0 \\ P_{c1} - C_1\lambda^{-1}f_{c1} = 0 \\ P_{R1} - R_1 f_{R1} = 0 \\ P_{R12} - R_{12}f_{R12} = 0 \\ P_{c2} - C_2\lambda^{-1}f_{c2} = 0 \\ P_{c3} - C_3\lambda^{-1}f_{c3} = 0 \\ P_{R2} - R_2 f_{R2} = 0 \\ p_{R23} - R_{23}f_{R23} = 0 \end{cases} \quad (9)$$

where λ represents the time differentiation operator and $\lambda^{-1}$ indicates integration over time. From Eq. (9.1), i.e. $\alpha_1(-P_{c1} + P_{R12} + P_{C2}) + (1-\alpha_1)f_{R12} = 0$, in case the water in tank 1 or tank 2 reach the level 0.5 the pipe $R_{12}$ become conducting, α₁ = 1, and $P_{c1} - P_{c2} = P_{R12}$, else, $\alpha_1 = 0$, and $f_{R12} = 0$. The TCM for this system of equations contains the relations between each of the variables. For example, Eq. (9.6) embodies a temporal relation between $P_{c1}$ and $f_{c1}$ and Eq. (9.2) a relation between $P_{c2}$, $P_{c3}$ and $P_{R23}$ that is only active when α₂= 1.

In TRANSCEND framework, only the three values −, 0, + are used to indicate values that are too low, normal, and too high, with respect to some nominal value, respectively. For example, a value of a model variable that is measured to be above its nominal value is marked +. In case the outflow through pipe line $R_{12}$ of the tank system in Fig. 1 is too high, this is represented by $f_{R12}^+$. To find parameter deviations, a back propagation algorithm is used in TRANSCEND framework [27]. In qualitative matrix algebra this is equivalent to repeated multiplication of the initial deviation with the transpose TCM. Next, predictions of future system behavior are generated for each of the possible parameter deviations. To achieve a suffiently high order prediction for the measured variable the initial deviation is repeatedly multiplied with the TCM. The TCM is derived from an implicit model formulation (9) that includes mode selection parameters, $\alpha_i$, to switch between equations. The possible causal assignments of higher relations are then made mutually exclusive by introducing selection parameters [27]. The TCG for three thank plant is shown in Fig. 10. Junctions and resistors define instantaneous magnitude relations, and capacitors and inertias define temporal effects on causal edges. For this example, to simplify the TCG structure, all =links and corresponding variables have been removed.

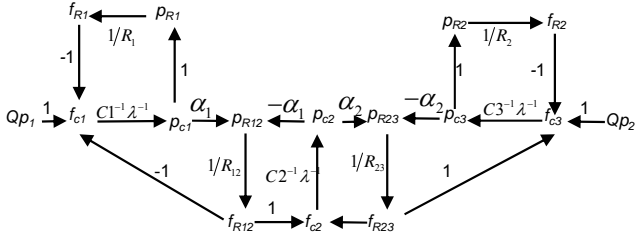

Fig. 10. Temporal Causal Graph of Three Tank System.

The resulting TCM is given by:

$$\begin{bmatrix} 1 & 0 & 0 & -1 & 0 & -1 & 0 & 0 & 0 & 0 & 0 & 0 & 0 \\ \lambda^{-1} & 1 & 0 & 0 & 0 & 0 & 0 & 0 & 0 & 0 & 0 & 0 & 0 \\ 0 & 1 & 1 & 0 & 0 & 0 & 0 & 0 & 0 & 0 & 0 & 0 & 0 \\ 0 & 0 & 1 & 1 & 0 & 0 & 0 & 0 & 0 & 0 & 0 & 0 & 0 \\ 0 & \alpha_1 & 0 & 0 & 1 & 1 & 0 & -\alpha_1 & 0 & 0 & 0 & 0 & 0 \\ 0 & 0 & 0 & 0 & 1 & 1 & 0 & 0 & 0 & 0 & 0 & 0 & 0 \\ 0 & 0 & 0 & 0 & 0 & 1 & 1 & 0 & 0 & -1 & 0 & 0 & 0 \\ 0 & 0 & 0 & 0 & 0 & 0 & \lambda^{-1} & 1 & 0 & 0 & 0 & 0 & 0 \\ 0 & 0 & 0 & 0 & 0 & 0 & \alpha_2 & 1 & 1 & 0 & -\alpha_2 & 0 & 0 \\ 0 & 0 & 0 & 0 & 0 & 0 & 0 & 0 & 1 & 1 & 0 & 0 & 0 \\ 0 & 0 & 0 & 0 & 0 & 0 & 0 & 0 & 0 & 1 & 1 & 0 & -1 \\ 0 & 0 & 0 & 0 & 0 & 0 & 0 & 0 & 0 & 0 & \lambda^{-1} & 1 & 0 \\ 0 & 0 & 0 & 0 & 0 & 0 & 0 & 0 & 0 & 0 & 0 & 1 & 1 \\ 0 & 0 & 0 & 0 & 0 & 0 & 0 & 0 & 0 & 0 & 0 & 1 & 1 \end{bmatrix} \begin{bmatrix} f_{c1} \\ P_{c1} \\ P_{R1} \\ f_{R1} \\ P_{R12} \\ f_{R12} \\ f_{c2} \\ P_{c2} \\ P_{R23} \\ f_{R23} \\ f_{c3} \\ P_{c3} \\ P_{R2} \\ f_{R2} \end{bmatrix} \quad (10)$$

The predictions of the TCM are parametrized by the active mode. This leads to more efficient diagnosis compared to the use of a bank of TCMs, which, in this case of two switches, would consist of four TCMs that need to be processed separately. The fault detectability and fault isolability of each parameter is gained from the $\{D_b, I_b\}$ values of the four FSMs. This information can be summarized in tables 4, 5, 6, 7 and 8.

Table 4: FSM at mode 1 ($\alpha_1 = \alpha_2 = 1$)

| Mode 1 | $P_{c1}$ | $P_{c2}$ | $P_{c3}$ | $D_b$ | $I_b$ |
|---|---|---|---|---|---|
| $R_1^+$ | 0+ | 00 | 00 | 1 | 1 |
| $C_1^-$ | +- | 0+ | 00 | 1 | 1 |
| $R_{12}^+$ | 0+ | 0- | 00 | 1 | 1 |
| $C_2^-$ | 0+ | +- | 0+ | 1 | 1 |
| $R_{23}^+$ | 00 | 0+ | 0- | 1 | 1 |
| $C_3^-$ | 00 | 0+ | +- | 1 | 1 |
| $R_2^+$ | 00 | 00 | 0+ | 1 | 1 |

Table 5: FSM at mode 2 ($\alpha_1 = 0, \alpha_2 = 1$)

| Mode 2 | $P_{c1}$ | $P_{c2}$ | $P_{c3}$ | $D_b$ | $I_b$ |
|---|---|---|---|---|---|
| $R_1^+$ | 0+ | 00 | 00 | 1 | 1 |
| $C_1^-$ | +- | 0+ | 00 | 1 | 1 |
| $R_{12}^+$ | 0+ | 0- | 00 | 1 | 1 |
| $C_2^-$ | 0+ | +- | 00 | 1 | 1 |
| $R_{23}^+$ | 00 | 00 | 00 | 0 | 0 |
| $C_3^-$ | 00 | 00 | +- | 1 | 1 |
| $R_2^+$ | 00 | 00 | 0+ | 1 | 1 |

Table 6: FSM at mode 3 ($\alpha_1 = 0, \alpha_2 = 1$)

| Mode 3 | $P_{c1}$ | $P_{c2}$ | $P_{c3}$ | $D_b$ | $I_b$ |
|---|---|---|---|---|---|
| $R_1^+$ | 0+ | 00 | 00 | 1 | 1 |
| $C_1^-$ | +- | 00 | 00 | 1 | 1 |
| $R_{12}^+$ | 00 | 00 | 00 | 0 | 0 |
| $C_2^-$ | 00 | +- | 0+ | 1 | 1 |
| $R_{23}^+$ | 00 | 0+ | 0- | 1 | 1 |
| $C_3^-$ | 00 | 0+ | +- | 1 | 1 |
| $R_2^+$ | 00 | 00 | 0+ | 1 | 1 |

Table 7: FSM at mode 4 ($\alpha_1 = \alpha_2 = 1$)

| Mode 4 | $P_{c1}$ | $P_{c2}$ | $P_{c3}$ | $D_b$ | $I_b$ |
|---|---|---|---|---|---|
| $R_1^+$ | 0+ | 00 | 00 | 1 | 1 |
| $C_1^-$ | +- | 00 | 00 | 1 | 1 |
| $R_{12}^+$ | 00 | 00 | 00 | 0 | 0 |
| $C_2^-$ | 00 | +0 | 00 | 1 | 1 |
| $R_{23}^+$ | 00 | 00 | 00 | 0 | 0 |
| $C_3^-$ | 00 | 00 | +- | 1 | 1 |
| $R_2^+$ | 00 | 00 | 0+ | 1 | 1 |

Table 8: Fault Detectability and fault Isolability of components

| $\theta$ | Detectability | Isolability |
|---|---|---|
| $R_1^+$ | all-mode | all-mode |
| $C_1^-$ | all-mode | all-mode |
| $R_{12}^+$ | mode 1, 2 | mode 1, 2 |
| $C_2^-$ | all-mode | all-mode |
| $R_{23}^+$ | mode 1, 3 | mode 1, 3 |
| $C_3^-$ | all-mode | all-mode |
| $R_2^+$ | all-mode | all-mode |

## 4. Conclusions and discussion

In this paper, causal properties of bond graphs are used to generate the elimination schemes such that direct and deduced redundancies can be expressed only in terms of known process variables. First, a quantitative fault diagnosis framework for hybrid systems is developed basing on HBG and on a set of unified constraint relations called global analytical redundancy relations (GARRs). These relations can be derived systematically from the diagnostic hybrid bond graph (DHBG). The GARRs explicitly show the system component fault detectability and fault isolability and generate alarm signals for effective and efficient fault detection and isolation (FDI). The quantitative method treats sensor, actuator and parameter faults which are of three types abrupt, progressive and intermittent. Noise and robustness issues are considered in such method. The application of the quantitative diagnosis is constrained to both open and closed loop systems. However, the subset of the equations of complex models with implicit relations, complex non-linearities, and algebraic loops, etc., cannot be resolved.

In addition to the quantitative method, another qualitative approach dealing with temporal causal graph (TCG) is used. This method allows one to ameliorate fault isolation, it treats only faults parameter indeed, noise and robustness issues are not considered in such diagnosis and only abrupt faults are handled. The application of this method is constrained to only open-loop systems because it relies on temporal trends of systems evolution obtained from the measurements which may not show any deviation if they are controlled. It is well known that controllers try to hide the fault effects. The qualitative approach overcomes limitations of quantitative schemes, such as convergence and accuracy problems in dealing with complex non-linearities and lack of precision of parameter values in system models. The qualitative reasoning scheme is fast, but it has limited discriminatory ability.

As presented in this article, a single approach for diagnosis has limitations, and it does not satisfy all the requirements to perform a good diagnosis. Hence, several works can be found in literatures that combine diagnostic approaches. The objective is to find two or several approaches that complete each other, in a way that the qualities of one approach can overcome the drawback of another. In a future work, we will focus on complex models with implicit relations, complex non-linearities, and algebraic loops.